\documentstyle[prl,aps,twocolumn,epsf]{revtex}

\begin{document} 

\title{Nonequilibrium Precursor Model for the Onset of Percolation in a 
Two-Phase System}

\author{Marcelo Gleiser$^{1}$, Rafael Howell$^{1}$ and Rudnei O. Ramos$^{1,2}$}

\address{
{\it $^{1}$Department of Physics and Astronomy, Dartmouth College,}\\
{\it  Hanover, New Hampshire 03755-3528, USA}\\
{\it $^{2}$Departamento de F\'{\i}sica Te\'orica,}
{\it Instituto de F\'{\i}sica, Universidade do Estado do Rio de Janeiro,}\\
{\it 20550-013 Rio de Janeiro, RJ, Brazil}}

\maketitle

\begin{abstract} 

Using a Boltzmann-like equation, we investigate the nonequilibrium dynamics of 
nonperturbative fluctuations 
within the context of Ginzburg-Landau models. 
As an illustration, we examine how a two-phase system initially prepared in
a homogeneous, low-temperature phase becomes populated by 
precursors of the
opposite phase as the temperature is increased. We
compute the critical value of the order parameter for the onset
of percolation, which signals the breakdown of the conventional dilute gas
approximation. 

\noindent 
PACS number(s): 05.70.Ln, 05.70.Fh, 64.60.Ak

\end{abstract} 

\vspace{0.34cm}

The process by which an ordered low-temperature system approaches disorder as 
its temperature is increased has far-ranging applications in many areas of 
physics. In systems which allow for low-temperature
symmetry breaking, such processes would
describe symmetry restoration.
There is considerable overlap between treatments found in the condensed
matter literature within the context of Ginzburg-Landau models 
\cite{Gunton,Langer}
and those found in the high-energy physics literature through the use of
temperature-corrected effective potentials \cite{Lebelac}, although clearly
there are several crucial differences as well \cite{Goldenfeld}. In this
letter, we would like to explore an issue which is of interest to both areas,
namely, how to describe the dynamics of nonperturbative
thermal fluctuations in simple
systems, modeled by the Ginzburg-Landau model.   

It is well-known that ferromagnets will become paramagnetic above a certain
critical temperature. It is also well-known that such emergence of
disorder is due to the nucleation of ferromagnetic domains of the opposite
phase \cite{Magnets}. The dynamics of the domain interfaces, as well as 
their growth, is of interest in many diverse areas, from materials science to
particle physics to cosmology, even if some systems 
require more complicated order parameters.
Examples include the recent experiments on
Bose-Einstein condensation in dilute atomic gases \cite{bosereview}
and the study of the growth of the condensate \cite{BFR}; 
ultra-relativistic heavy-ion
collision experiments \cite{DCC}; 
formation
of topological defects both in the laboratory, via pressure quench experiments
\cite{Zurek}, or in the early Universe \cite{Vilenkin}; and
the nematic-isotropic
transition in liquid crystals \cite{de Gennes}. Diverse as these systems are,
they all have one feature in common; their change in behavior is due to the
onset of nonequilibrium conditions, which are poorly understood.

In studying these kinds of problems one usually starts with mean-field
theory, or some of its microscopic versions, such as the 
equilibrium one-loop 
approximation in field theory or the Hartree 
approximation. In these approximations, localized, high amplitude fluctuations
are neglected and replaced by an average interaction of the system with
the thermal environment \cite{Goldenfeld}. However, 
this approximation
breaks down as the system approaches criticality
and these fluctuations become more important.

It has long been recognized \cite{landau} that in order to fully understand
the dynamics of a given system one must go beyond the mean field approximation.
One approach is to invoke time-dependent renormalization group techniques
\cite{Goldenfeld}.
Here, we would like to propose an alternative approach, based on the  
dynamics of large-amplitude fluctuations, from which we can
examine the nonequilibrium properties of the system. As we will show, our
approach is valid up to the onset
of percolation, which is known to occur before criticality 
for 3-dimensional systems.

Let us start by considering a standard Ginzburg-Landau model where
local fluctuations about homogeneous equilibrium have the
free energy

\begin{equation}
F(\phi,T) = \int d^3x \left[\frac{b}{2} |\nabla \phi|^2 + V(\phi)\right]\;,
\label{ham}
\end{equation}

\noindent
with $V(\phi) = V_0 + a (\theta-1) \phi^2/2
+ \lambda \phi^4/4$, 
where $a,\;b,\;V_0$ and $\lambda$
are (positive) constants and $\theta$ is the temperature ratio
$T/T_c$. {}For convenience, we have 
added the constant term $V_0$ 
to fix the minima of the free energy density
at $T<T_c$ at zero, $\langle \phi \rangle = \pm \sqrt{a(1-\theta)/\lambda}
\equiv \phi_\pm$, which then gives that
$V_0=a^2(1-\theta)^2/(4\lambda)$. The constants $a$ and $b$ 
can easily be scaled away 
and $F(\phi,T)$ can be made dependent only on the temperature ratio
$\theta$ and on the coupling constant $\lambda$.

We choose to study the dynamics of the fluctuations as the system is heated
from its $T=0$ state, where it is in
one of its ordered states, say $\phi_-$, to a temperature $T<T_c$,
focusing on its evolution to a final equilibrium state determined by a
time-independent
value of the order parameter $\langle \phi \rangle$ at temperature $T$. 
We model the fluctuations away from the initial equilibrium state
as Gaussian shaped, spherically-symmetric configurations
with a core value $\phi_C$ and radius $R$. 
These precursors are also called subcritical
bubbles and treatments involving them have been successfully used in many
other contexts before \cite{subcritical}. By expressing the amplitude of the
fluctuation as $\phi_A = \phi_C - \phi_-$, the fluctuations are parameterized 
as $\phi_f (r) = \phi_A \; \exp(-r^2/R^2) + \phi_-$.
[The Gaussian satisfies the physical boundary conditions -- regularity
at the origin and asymptotic approach to the background, while costing very
little free energy. See Ref. \cite{subcritical} for details.]

Since we are interested in fluctuations which can probe the other available
free-energy minimum,
their amplitudes can 
be easily determined by the condition that $\phi_f (r)$ represents
unstable fluctuations inside the $(-)$-vacuum phase.
We then simply have, from  the symmetric double well potential
used in Eq. (\ref{ham}) (note that
small-amplitude fluctuations are already incorporated in the mean-field 
approximation to the free-energy density), that those fluctuations with 
$\phi_C \geq \phi_{\rm max}=0$ are the ones probing the $(+)$ phase.
As for $R$, we allow for fluctuations larger 
than the correlation length, $R_{\rm min} = \xi(T)$,
where $\xi(T)= [V''(\phi_\pm)]^{-1/2}$, consistent with the natural 
coarse-graining scale dictated by the continuous free energy. 
Substituting $\phi_f (r)$ in (\ref{ham}) we obtain the 
free energy barrier for a fluctuation with amplitude $\phi_A$ and radius $R$
as $F_f (R,\phi_A,T) = (3 \sqrt{2}\pi^{3/2} \phi_A^2/8)
R + \pi^{3/2} \phi_A^2 [\sqrt{2} V''(\phi_-)/8+
\sqrt{3} V'''(\phi_-) \phi_A/54 + V^{(4)}(\phi_-) \phi_A^2/192] R^3$. We will
refer to the fluctuations with $\phi_C \geq \phi_{\rm max}$ as ``(+)-phase
fluctuations'' and the background phase as the ``($-$)-phase''.

We next study the dynamical evolution of these fluctuations. 
{}For this we use the Boltzmann-like equation derived in Ref. \cite{GHK} for
the distribution function of (+)-phase
fluctuations of radius $R$ and amplitude $\phi_A$,
$f_+(R,\phi_A,t)$, which satisfies the equation

\begin{equation}
\frac{\partial f_+}{\partial t} =  |v| 
\frac{\partial f_+}{\partial R}
+ (1-\gamma) G_{-\to +} - \gamma G_{+\to -}\;,
\label{boltz}
\end{equation}

\noindent
where the first term in the rhs incorporates the collapse of subcritical
domains, which we approximate as having constant velocity
$v= \partial R/\partial t$. In a forthcoming publication we will show that 
this is a valid approximation for most of the interesting range of 
bubble radii. The second term describes the
thermal nucleation of fluctuations of the $(+)$-phase inside the $(-)$-phase,
with nucleation rate $G_{-\to +}$, while the last term
describes the inverse process, with rate $G_{+\to -}$.
{}For a degenerate potential these two rates are the same, which
we express in terms of the free energy of the fluctuations, $F_f$, as
given by a standard Gibbs distribution: $G(\phi_A,R)
\equiv G_{-\to +} = G_{+\to -}
=A T^4 \exp(-F_f/T)$, where $A$ is a constant. 
Note also that from Eq. (\ref{boltz}),
detailed balance imposes that the ratio $A/|v|$ be constant, which will 
be taken as a free parameter in our model; it can be 
determined for specific 
models, as shown in Ref. \cite{GHK}. In fact,
the ratio $A/|v|$ must
depend on dynamical quantities which
are, in principle, expressable in terms of the only two parameters in the
model free energy, $\lambda$ and $\theta$, which must control heat diffusion 
and fluctuations
dynamics and can be mapped to specific applications.
$\gamma$ in Eq. (\ref{boltz}) is the total fraction of volume in
the $(+)$-phase, defined by \cite{GHK}

\begin{equation}
\gamma = \int_{\phi_{\rm max}}^{+\infty}d\phi_C \int_{\xi(T)}^{+\infty}
d R \; \frac{4 \pi R^3}{3}\; f_+(R,\phi_A,t)\;.
\label{fraction}
\end{equation}

\noindent
Note that from our initial condition at $t=0$, we have
$\gamma(t=0)=0$ and, in the asymptotic equilibrium regime at temperature 
$0< T \leq T_c$, $0 < \gamma_{\rm eq} \leq 1/2$.
Eq. (\ref{boltz}), together with (\ref{fraction}), 
is an integro-differential equation that we numerically solve for $\gamma$.
The result is shown in {}Fig. 1 for different values of temperature and
$\lambda=1$.

\begin{figure}[hc]
\epsfysize=6 cm 
{\centerline{\epsfbox{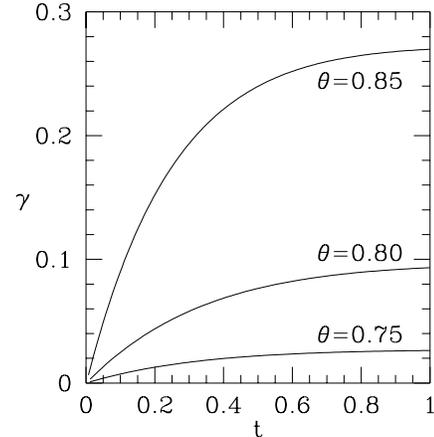}}}
\caption{The volume fraction $\gamma$ as a function of time for $\lambda=1$
and $A/|v|=1$.}
\end{figure}

$\gamma_{\rm eq}$ can be computed 
from Eq. (\ref{boltz}) by setting the time derivative term in the lhs 
to zero. The resulting expression for $\gamma_{\rm eq}$, using
Eq. (\ref{fraction}), is $\gamma_{\rm eq} = I/(1+2 I)$,
where

\begin{equation}
I= \!\!\int_{\phi_{\rm max}}^{+\infty}\!\!\! d\phi_C \int_{\xi(T)}^{+\infty}
d R \; \frac{4 \pi R^3}{3} \int_R^{+\infty} dR' 
\frac{A}{|v|}T^4 e^{-F_f/T}\;.
\label{I}
\end{equation}

\noindent
Writing $\gamma(t) \equiv \gamma_{\rm eq} {\cal B}(t)$, 
and using Eqs. (\ref{boltz}) and
(\ref{fraction}), we find an equation for ${\cal B}(t)$, after
integrating all terms in Eq. (\ref{boltz}) by $\int_{\phi_{\rm max}}^{+\infty} 
d\phi_C \int_{\xi(T)}^{+\infty} dR \; (4\pi R^3/3)$,

\begin{equation}
\frac{d {\cal B}(t)}{dt} + \frac{\Gamma}{\gamma_{\rm eq}} {\cal B}(t)
-\frac{\Gamma}{\gamma_{\rm eq}} = 0 \;,
\label{diff}
\end{equation}

\noindent
where $\Gamma$, the total volume-integrated nucleation rate, is given by

\begin{equation}
\Gamma = \int_{\phi_{\rm max}}^{+\infty} 
d\phi_C \int_{\xi(T)}^{+\infty} dR \frac{4\pi R^3}{3} G(\phi_C,R)\;.
\label{G3}
\end{equation}

\noindent
The differential equation (\ref{diff}) has a simple solution, given by
${\cal B}(t) = 1-\exp(-t/\tau)$, where $\tau=\gamma_{\rm eq}/\Gamma$ 
is the equilibration time-scale. Therefore, the analytical solution 
for $\gamma(t)$ is
$\gamma(t) = \gamma_{\rm eq} \left(1-e^{-t/\tau}\right)$.
This solution fits very well the numerical solution for $\gamma$ shown in
{}Fig. 1. In {}Fig. 2 we compare the theoretical and numerical results for 
the equilibration time-scale $\tau$, as a function of the temperature, 
for two different values of $\lambda$; the agreement is quite striking.
The results in {}Fig. 2 reveal a peak in the equilibration time-scale.
{}For small temperatures, the equilibration time grows continuously until 
it reaches a maximum at the temperature $T_{\rm max}(\lambda)$. This is 
reminiscent of the critical slowing down behavior characteristic of critical 
phenomena, although we cannot recover the discontinuity at the critical 
point with our simple model. What we do provide is a dynamical picture of 
the approach to criticality, which we expand on below. As the
temperature increases, a larger fraction of the volume of the initial
state in the $(-)$-phase is populated by 
fluctuations of the (+)-phase; also, 
as the free-energy barrier decreases with increasing temperature,
these fluctuations will become larger
in size.
Hence, their equilibration time-scales grow with increasing $T$, as 
displayed in {}Fig. 2. 
{}For temperatures larger than $T_{\rm max}$, the 
equilibration time decreases, vanishing at the mean-field critical
temperature $T_c$.  
This is due to the fact that at 
$T > T_{\rm max}$ the system is described by a free-energy density
with a single minimum at $\phi=0$; thus, 
the true critical temperature is not at $T_c$.
This is in accordance with what is expected
when corrections beyond the mean field 
to the potential are taken into account \cite{bellac} and explicitly
seen in large-scale
Langevin simulations performed on the lattice \cite{BG}.
{}From {}Fig. 2 our model predicts 
the values $T_{\rm max} \simeq 0.79\; T_c$ for
$\lambda=1$ and $T_{\rm max} \simeq 0.97\; T_c$ for
$\lambda=0.1$. These results are dependent on the ratio $A/|v|$
which entails the microscopic details of a given model under study.
Physical lower bounds on this ratio will be discussed below.

\begin{figure}[hc]
\epsfysize=6 cm 
{\centerline{\epsfbox{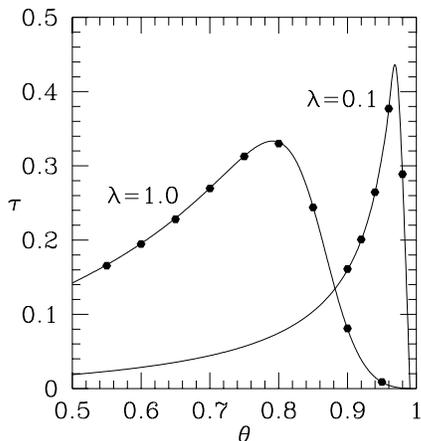}}}
\caption{The equilibration time $\tau$ for $\lambda=1$ and $\lambda=0.1$,
with $A/|v|=1$.
The dots are the numerical results from Eq. (\ref{boltz})
and the lines the theoretical result, $\tau_{\rm theor}
=\gamma_{\rm eq}/\Gamma$.}
\end{figure}

As the temperature is increased and the volume-density 
of (+)-phase fluctuations grows, the system will eventually
reach an instability point beyond which domains of the (+)-phase will
grow by percolating with their nearest neighbors \cite{stauffer}. 
The question is at what
temperature such percolating instability occurs. In order to answer this
question, we take full advantage of our dynamical model. [A preliminary
approach can be found in Ref. \cite{GHK}.] Since
correlation-volume fluctuations have the smallest free-energy barrier, they
will be statistically dominant.
In order to model the percolation-instability,
consider a domain of the (+)-phase of correlation-length radius
$\xi$. 
There are three main processes that can 
change its volume: shrinking due to its surface tension;
growth due to the thermal nucleation of another (+)-phase domain of radius
$R$
just outside it,
which accounts for a change of volume
$\Delta V = 4 \pi [(\xi + 2 R)^3-\xi^3]/3$; and thermal destruction
of the correlation size fluctuation due to inverse nucleation,
that changes the volume by $\Delta V'= 4 \pi \xi^3/3$. We thus arrive at
an approximate equation describing the rate of change of $V_\xi$:

\begin{eqnarray}
\frac{d V_\xi}{d t} &\simeq &  - |v| 4 \pi \xi^2 \!\!+\!\!    
\int_{\phi_{\rm max}}^{+\infty} \!
d\phi_C \! \int_{\xi(T)}^{+\infty} \! dR \frac{4\pi R^3}{3} G(\phi_C,R)
\Delta V \nonumber \\
&-& \int_{\phi_{\rm max}}^{+\infty} \!
d\phi_C \! \frac{4 \pi \xi^4}{3} G(\phi_C,\xi) \Delta V' \equiv 
4\pi\xi^2 v_{\rm eff}
\;.
\label{volume}
\end{eqnarray}

In {}Fig. 3 we show the numerical solution for the effective velocity,
$v_{\rm eff}$, as a function of temperature.
The temperature for which $v_{\rm eff} > 0$, $T_{\rm perc}(\lambda)$,
indicates the onset 
of percolation for correlation-volume fluctuations. 

\begin{figure}[hc]
\epsfysize=6 cm 
{\centerline{\epsfbox{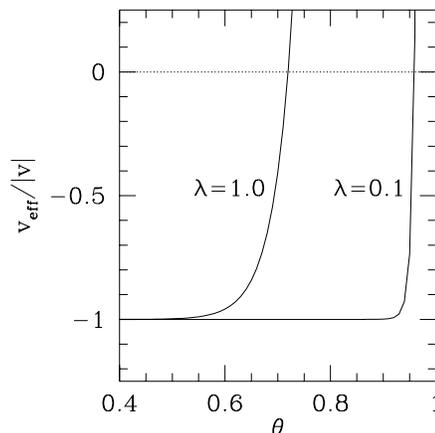}}}
\caption{The effective velocity of correlation length fluctuations,
$v_{\rm eff}=(4\pi\xi^2)^{-1}dV_\xi/dt$ as a function of temperature for 
$\lambda=1$ and $\lambda=0.1$. $A/|v|$ is again set to one.}
\end{figure}

\noindent
Our numerical
results give the values $T_{\rm perc} \simeq 0.72\; T_c$ for $\lambda=1$
and $T_{\rm perc} \simeq 0.96\; T_c$ for $\lambda=0.1$.
$T_{\rm perc}$ obtained
with equation (\ref{volume}) is very close to $T_{\rm max}$
given before. As percolation sets in, 
Eq. (\ref{boltz}) begins to underestimate the growth of 
fluctuations and the further development of 
the system. $T_{\rm perc}$ then determines the limit of validity
of Eq. (\ref{boltz}), or the breakdown of the dilute gas approximation,
beyond which coalescence of phase fluctuations begins to be of 
importance. Nevertheless, Eqs. (\ref{boltz}) and (\ref{volume})
describe quite well the dynamics until the
onset of continuous percolation as well as the equilibrium 
properties of the system. (Note that continuous percolation, as opposed
to lattice percolation, is very poorly
understood, and only within simple 2-dimensional mathematical models,
such as the Boolean-Poisson model \cite{stauffer}.) 

{}Finally, it is important to test the validity of this model with respect
to the calculation of the free energy of the fluctuations.  
It is clear that as the free energy $F_f$ for fluctuations drops below
$k_B T$ we no longer can distinguish them from simple thermal noise,
which then becomes statistically dominant; the description of the 
nucleation of large-amplitude fluctuations with rate $G$
becomes meaningless. Using the temperatures $T_{\rm max}$ and 
$T_{\rm perc}$, we can set a lower bound on the value of 
$A/|v|$.
This is shown in {}Fig. 4 for the case
of $\lambda =1$, where we show how 
$T_{\rm max}$ and $T_{\rm perc}$ change with $A/|v|$. The condition
$F_f/T|_{T_n} = 1$ applied to the fluctuations of lowest
free energy,  $\phi = \phi_{\rm max}$ and $R=\xi$, gives $T_n \gtrsim
0.87\; T_c$ for $\lambda=1$ and $T_n \gtrsim
0.99\; T_c$ for $\lambda=0.1$.                     

\begin{figure}[hc]
\epsfysize=6 cm 
{\centerline{\epsfbox{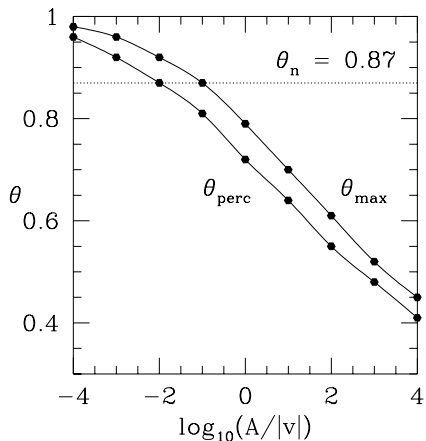}}}
\caption{$\theta_{\rm max} = T_{\rm max}/T_c$ and $\theta_{\rm perc}=
T_{\rm perc}/T_c$ as a function of $A/|v|$, for $\lambda=1$. The dotted line
is defined by the condition $F_f/T|_{T_n} = 1$, from which we obtain the bound 
$A/|v| > 10^{-2}$.}
\end{figure}

Summarizing, we have presented a simple model based on
the dynamics of phase fluctuations that is able to provide 
a dynamical description of how a continuous ordered system described by a 
Ginzburg-Landau free energy approaches its percolation
instability. The model allows us to compute the 
temperature for the onset
of percolation, which signals the breakdown of the conventional dilute gas
approximation, offering also a novel way to
estimate the actual critical temperature in a Ginzburg-Landau system, 
that can be
read from the results of {}Fig. 2. {}Furthermore, the model studied here,
despite its simplicity, exhibits a dynamical picture
of symmetry restoration and the breakdown
of mean-field theory observed both numerically and analytically, without 
recourse 
to large-scale numerical simulations.
The model can easily be extended to different
systems including inhomogeneous nucleation, or systems outside the Ising
universality class.
We expect to report soon on these applications. 

We thank J. D. Gunton and D. Huse for discussions.
M. G. is supported by NSF grants PHY-0070554 and
PHY-9453431.
R.O.R. was supported by Conselho Nacional de
Desenvolvimento Cient\'{\i}fico e Tecnol\'ogico - CNPq (Brazil)
and SR-2 (UERJ) and by the ``Mr. Tompkins Fund for Cosmology and Field Theory''
at Dartmouth.

\end{document}